\documentclass[
amsmath,amssymb,
prfluids,
]{revtex4-2}

\usepackage{graphicx}
\usepackage{dcolumn}
\usepackage{bm}
\usepackage[colorlinks,urlcolor=blue,linkcolor=blue,citecolor=blue]{hyperref}

\usepackage{graphicx}
\usepackage{dcolumn}
\usepackage{bm}
\usepackage[mathlines]{lineno}
\usepackage{orcidlink}
\usepackage{accents}
\usepackage{xcolor}
\usepackage{empheq}

\definecolor{Gold}{RGB}{255, 215, 0}

\newcommand{\mdiamond}{%
  \mathbin{%
    \scalebox{1.}[1.6]{$\diamond$}
  }%
}

\newcommand{\mcirc}{%
  \mathbin{%
    \scalebox{1.6}{$\circ$}
  }%
}

\newcommand{\mplus}{%
  \mathbin{%
    \scalebox{1.2}{$+$}
  }%
}

\newcommand{\vect}[1]{\boldsymbol{#1}}
\newcommand{\mathsbf}[1]{{\boldsymbol{\mathsf{#1}}}}
\newcommand*{\dt}[1]{%
\accentset{\mbox{\large\bfseries .}}{#1}}

\begin{document}
\title{Conversions between kinetic and surface energy in periodically forced multiphase turbulence}

\author{J. Vahé\orcidlink{0009-0005-5947-7860}}
\affiliation{CNRS, CORIA, UMR 6614, Normandy Univ., UNIROUEN, INSA Rouen, 76000 Rouen, France}

\author{F. Thiesset\orcidlink{0000-0001-7303-5106}}
\email{fabien.thiesset@cnrs.fr}
\affiliation{CNRS, CORIA, UMR 6614, Normandy Univ., UNIROUEN, INSA Rouen, 76000 Rouen, France}

\begin{abstract}
    In multiphase flows, kinetic and interfacial energies coexist, and their mutual conversion can potentially influence the overall energy balance. However, in statistically steady flows these energy reservoirs remain constant, making such conversions undetectable. For them to be observed, a degree of unsteadiness must be introduced, here provided by the deliberate use of a fluctuating time-periodic input of kinetic energy into the system. The main focus of the present work is on the dynamical cycle connecting energy injection, conversion, and dissipation which we explore using {direct} numerical simulations of multiphase homogeneous isotropic turbulence, subjected to periodic forcing. The database includes various Reynolds and Weber numbers and volume fractions in the dense regime. To interpret and replicate the observed dynamics, we reformulate the \textit{Ka-Pi-bara} model of \cite{Bos2026} (an extension of the $k$--$\epsilon$ model) in terms of total energy (the sum of kinetic and surface energy), which we further enhance by adding equations for the surface energy and its destruction. This model accurately captures a key feature of turbulence: non-equilibrium effects, seen as the phase lag between kinetic energy and its rate of dissipation, which are found to operate also in multiphase flows. Linearizing the model highlights the various relevant time scales of the system and provides predictions of how different observables are coupled and respond to the energy input. 
\end{abstract}
\maketitle

\section{Introduction}

Flows of immiscible fluids are common in both natural and industrial systems. In many cases, the interfacial surface area between the two phases is a critical parameter, controlling phenomena such as heat and mass transfer. Yet, predicting the interface surface area requires understanding the complex coupling between the flow and the interface which remains a major challenge. This interplay becomes clear when considering energy balances: the interface stores surface energy through surface tension, coexisting with kinetic energy \cite{Kataoka1986,Dodd2016,VelaMartin2021,Calado2024}. A key open question is how these different forms of energy interact.

Recent studies \cite{Mukherjee2019,Freund2019,Pandey2020,CrialesiEsposito2022,Pan2022,Cannon2024,Ramirez2024,Narula2026,Thiesset2025} have gradually converged toward a coherent framework for describing energy exchanges in multiphase turbulence. In this picture, a portion of the kinetic energy injected at large scales cascades to smaller scales, much like in single-phase flows. Concurrently, when surface tension is present, part of the injected energy is also converted into interfacial or surface energy as interfaces become progressively distorted and fluid structures break up. At smaller scales, interfaces may relax, fluid structures could also coalesce, converting surface energy back into kinetic energy. Both the turbulent cascade and energy conversion processes continue down to the smallest scales, where kinetic energy is ultimately dissipated into heat.

Multiphase flows are thus characterized by a complex interplay of energy injection, cascade, conversion, and dissipation. At steady state, both kinetic and surface energy remain constant, and energy conversions vanish, reflecting an equilibrium between the two reservoirs and between energy injection and dissipation, as in single-phase flows. In turbulent flows, however, steady state can only be defined statistically, after averaging over long times. Instantaneously, spatial averages remain non-stationary, making turbulence a textbook example of an “out-of-equilibrium” system, where the injected energy requires finite time to cascade across scales before being dissipated \cite[see e.g.][]{Pearson2004,Cardesa2015,Goto2016}. Non-equilibrium effects are then manifested in particular in a time delay between fluctuations of kinetic energy and its rate of dissipation. 

Because in multiphase turbulence, energy conversion occurs alongside energy transfer, it can potentially lead to atypical out-of-equilibrium behaviours that remain largely unexplored. To date, only \citet{Mukherjee2019} have revealed an intriguing dynamical cycle linking kinetic energy injection, conversion into surface energy, and dissipation into heat. This study pursues three main objectives: (i) to analyse energy exchanges in unsteady multiphase flows, (ii) to examine the coupling and dissipation/destruction of kinetic and surface energies, and (iii) to clarify the out-of-equilibrium characteristics of multiphase turbulent flows for both forms of energy.

Here, we deliberately keep the flow in a non-stationary regime while retaining the possibility of meaningful statistical analysis. This is achieved by considering periodically forced turbulence, for which phase averaging can be employed in place of classical time averaging. Such a configuration has been widely used, in particular to investigate non-equilibrium effects \cite{Heydt2003,Bos2007} or to explore the role of intermittency of the large scales \cite{Chien2013,Bentkamp2025}. They also provide a convenient benchmark for one-point and two-point closure models \cite[see, e.g.,][and references therein]{Rubinstein2009,Yang2019,Fang2023,Bos2026}.

Periodically forced flows are known to exhibit two asymptotic responses. At very low forcing frequencies, the system has sufficient time to adjust to variations in energy input and therefore evolves through a sequence of locally steady states; this corresponds to the static limit. Conversely, at very high forcing frequencies, the turbulent cascade acts as a low-pass filter, so that dissipation does not respond to changes in the energy input, defining the frozen limit. While these extreme regimes are relatively straightforward to predict, more interesting dynamics arise at intermediate forcing frequencies, which is the regime of primary interest in the present study.

We conduct {direct} numerical simulations (DNS) of multiphase turbulence driven by harmonic forcing in a triply periodic domain using a standard Navier--Stokes solver coupled with an interface-capturing method. The resulting flow is homogeneous and isotropic. To interpret and reproduce the {DNS} observations, we propose a model that accounts for the coupling between kinetic and surface energies, as well as their respective dissipation or destruction mechanisms. This model builds upon a recent extension of the $k$--$\epsilon$ framework, referred to as the \textit{Ka-Pi-bara} model \cite{Bos2026}, which was developed to capture non-equilibrium effects in single-phase flows.

The extension to multiphase flows proceeds in two steps. First, we reformulate the model in terms of the total energy, defined as the sum of kinetic and surface energies, in order to incorporate the influence of interfaces on the kinetic energy dissipation rate. Second, we introduce additional evolution equations for the surface energy and its destruction, formulated in direct analogy with the equations governing kinetic energy and its dissipation rate in the $k$--$\epsilon$ model. Linearising the model allows us to predict the gain and phase lag of the different observables with respect to changes in the energy input.

The paper is organized as follows. First, the derivation of the model and the linearised solutions are presented in \S \ref{sec:model}. Details about the {DNS} procedure and the explored parameters are provided in \S\ref{sec:simulations}. Results are gathered and discussed in \S \ref{sec:results}. Section \ref{sec:conclusions} summarizes the main conclusions.
\section{Modelling} \label{sec:model}
\subsection{Single-phase flows}

The standard $k-\epsilon$ model \cite{Launder1974} for single phase, homogeneous, forced turbulence writes:
\begin{subequations}
    \begin{eqnarray}
        \dt{E}_k &=& F - \epsilon \label{eq:dot_Ek}\\
        \dt{\epsilon} &=& C_{1} F \frac{\epsilon}{E_k} - C_{2} \frac{\epsilon^2}{E_k} \label{eq:dot_eps},
    \end{eqnarray}
\end{subequations}
where $E_k = \frac{1}{2}\langle\rho |\vect{u}|^2\rangle$ is the kinetic energy, with the brackets denoting a volume average and $\rho$ is the density. The kinetic energy dissipation rate is defined by $\epsilon = \langle \mu ~\mathsbf{S}:\mathsbf{S}\rangle $ where $\mu = \nu/\rho$ is the dynamic viscosity, $\nu$ the kinematic viscosity and $\mathsbf{S}$ the strain rate tensor. The forcing is defined by $F = \langle \rho \vect{u}\cdot \vect{f}\rangle$ where $\vect{f}$ represents an acceleration field. Time derivatives of any quantity $X$ is denoted $\dt{X}$. {The flow we consider is statistically homogeneous, and therefore the $k-\epsilon$ model does not contain spatial derivatives.}

Equation \eqref{eq:dot_Ek} is exact while Eq. \eqref{eq:dot_eps} is a model equation for the time variations of the kinetic energy dissipation rate $\epsilon$. The latter depends on some model parameters $C_{1}$ and $C_{2}$ which needs to be determined. {For statistically homogeneous turbulence, the $k-\epsilon$ model is a set of coupled but closed ordinary differential equations (ODE) which provides estimates for the kinetic energy and dissipation, where the forcing $F$ is the input. In inhomogeneous flows, spatial derivatives needs to be included and the $k-\epsilon$ model becomes a set of partial differential equations.} The steady state solution of Eq. \eqref{eq:dot_Ek}, is $F^0=\epsilon^0$ (superscript 0 denote the steady state quantities). For Eq. \eqref{eq:dot_eps} to be compatible with the steady state solution $\dt{\epsilon}^0=0$, the model parameters $C_{1}$ and $C_{2}$ should be equal, i.e. $C_{1}$ = $C_{2} = C_k$. The $k-\epsilon$ model can thus be rewritten as:
\begin{subequations}
    \begin{eqnarray}
        \dt{E}_k &=& F-\epsilon \\
        \dt{\epsilon} &=& C_{k} \frac{\epsilon}{E_k} \dt{E}_k 
    \end{eqnarray} \label{eq:k_eps}
\end{subequations}
For turbulent flows with time-periodic forcing, the $k-\epsilon$ model predicts a finite time delay between forcing and dissipation, which depends on the ratio between the forcing frequency and the inverse of eddy turn-over time $C_k \epsilon^0 / E_k^0$. However, the model predicts that $E_k$ and $\epsilon$ are in phase, which is inconsistent with a wide range of observations supporting out-of-equilibrium dynamics.

In recent works by \citet{Bos2026,Bos2025}, a modified $k$–$\epsilon$ model was developed which properly reflects the time lag between $E_k$ and $\epsilon$. For this purpose, they started from the spectral balance equation for forced homogeneous turbulence. They assume that the large-scales kinetic energy follows $\dt{E}_k^< = F-\Pi$ while the small-scales kinetic energy evolves as $\dt{E}_k^> = \Pi-\epsilon$, where $\Pi$ is the kinetic energy transfer across scales. Then, using a Taylor-like expression for the energy transfer, i.e. $\Pi \sim {E_k^<}^{3/2}/L$, they obtained the so-called \textit{Ka-Pi-bara} model which writes (see eq. (59) of \cite{Bos2026}):
\begin{subequations}
    \begin{eqnarray}
        \dt{E}_k &=& F-\epsilon \\
        \dt{\epsilon} &=& C_{k} \frac{\epsilon}{E_k} \dt{E}_k  \left(1+\gamma \frac{F}{\epsilon}\right) - \gamma \dt{F}
    \end{eqnarray} \label{eq:kapibara}
\end{subequations}
In the system Eq. \eqref{eq:kapibara}, the symbol $\gamma = E_k^>/E_k^<$ represents the small-scale to large-scale kinetic energy ratio. The \textit{Ka-Pi-bara} assumes $\gamma$ is constant in time and appears as a second model parameter which needs to be determined. When $\gamma = 0$, the model reduces to the original $k$--$\epsilon$ formulation, showing that the delay between $E_k$ and $\epsilon$ is encoded into this constant $\gamma$.

\subsection{Multiphase flows}

Our purpose is now to extend this framework to multiphase flows where kinetic energy $E_k$ coexists with surface energy $E_s$ related to interfacial forces. For an interface with a constant surface tension coefficient $\sigma$, the surface energy is given by $E_s = \sigma A$ \cite{Mukherjee2019,VelaMartin2021,Calado2024}, where $A$ is the surface area density of the interface between the two phases. For statistically homogeneous flows, in absence of mass and heat exchanges, the evolution equation for the kinetic energy then writes \cite[e.g.][]{VelaMartin2021,Begemann2022,Calado2024,Boniou2025,Jain2025,Thiesset2025}:
\begin{eqnarray}
    \dt{E}_k = F - \epsilon - \dt{E}_s 
\end{eqnarray}
where $\dt{E}_s = \sigma \dt{A}$ corresponds to the conversion of kinetic energy into surface energy \cite{Dodd2016}. This implies that the evolution of the total energy $E_t = E_k+E_s$ is given by:
\begin{equation}
    \dt{E}_t = F-\epsilon \label{eq:dot_Et}.
\end{equation}
Equation \eqref{eq:dot_Et} indicates that the energy budget in multiphase turbulence is formally analogous to the one in single-phase turbulence, once the total energy $E_t$ is used in place of the kinetic energy $E_k$. It means that the conversion between kinetic and surface energy is conservative: surface energy has no intrinsic dissipation. All irreversible process occurs via viscosity: the total energy dissipation rate is the kinetic energy dissipation rate.

For modelling purposes, it thus seems natural to use $E_t$ instead of $E_k$ in the evolution equation for $\epsilon$ in the $k-\epsilon$ or \textit{Ka-Pi-bara} model, viz:
\begin{equation}
    \dt{\epsilon} = C_k \frac{\epsilon}{E_t} \dt{E}_t \left(1+\gamma \frac{F}{\epsilon}\right) - \gamma \dt{F}. \label{eq:dt_eps_Et}
\end{equation}
This equation can be derived using the same procedure as in \cite{Bos2025,Bos2026} assuming that the total energy $E_t$ can be used in place of the kinetic energy $E_k$ in the spectral balance equation, viz. $\gamma = E_t^>/E_t^<$ and assuming that $\Pi$ reads as the non-linear transfer term plus the scale-by-scale contribution due to the conversion between kinetic and surface energy; an argument confirmed by the theory of \cite{Pan2022}.

Several studies \cite[e.g.][among others]{Dodd2016,Mukherjee2019,Thiesset2025} have reported a relationship between kinetic energy dissipation and interfacial surface area. This effect enters via Equation~\eqref{eq:dt_eps_Et} which formally couples the equations governing $\epsilon$ and $E_s$. 

The evolution equation for the surface density $A$ is known \cite{Pope1988,Candel1990}. For homogeneous flows, it can be formulated in terms of $E_s$ as \cite{VelaMartin2021,Calado2024}:
\begin{eqnarray}
    \dt{E}_s = K E_s, \label{eq:stretch_rate}
\end{eqnarray}
where $K$ is the rate of production/destruction of surface area (or here surface energy). {This evolution for $E_s$ is exact. It reveals that the surface energy varies according to a linear source term $KE_s$. In homogeneous multiphase turbulence with no phase change, the stretch rate $K$ relates only to the velocity gradient tangent to the interface, i.e. $K = \langle \vect{\nabla}\vect{u}:\vect{n}\vect{n}\rangle_s$, where $\vect{\nabla}\vect{u}$ is the velocity gradient tensor, $\vect{n}$ is the normal vector to the interface, while $\langle \rangle_s$ denotes an area weighted average. Interface area increases when stretched ($K>0$) and decreases with compression ($K<0$) \cite{VelaMartin2021,Calado2024}. It also accounts for topological changes: breakup generally implies $E_s$ to increase and coalescence leads $E_s$ to decrease.} By analogy with $E_k$, we will write $K = P-\chi$ where $P$ relates to a surface energy production term associated with stretch and breakup, while $\chi$ is a surface energy destruction term due to compression and coalescence. This idea was already present in \cite{Vallet1999} and was reappraised in recent studies \cite{Mukherjee2019,Pandey2020,VelaMartin2021,CrialesiEsposito2022,Thiesset2025}. \citet{Vallet1999} suggested that $P$ is proportional to the inverse of the Kolmogorov time-scale, a postulate that is robustly verified in a variety of different situations including material surfaces \cite[e.g.][]{Yeung1990}, passive scalar iso-surfaces \cite[e.g.][]{Gauding2022}, premixed flames \cite[e.g.][]{Chu2024}, etc. Hence, we here assume $P = (\epsilon / \nu)^{1/2}$. 

However, much less is known about $\chi$. {The equation for the surface energy Eq. \eqref{eq:stretch_rate} indicates that the quantity $\epsilon_s = \chi E_s$ plays for $E_s$ the same role as $\epsilon$ for $E_k$. Similarly, the quantity $F_s = P E_s$ plays for $E_s$ the same role as $F$ for $E_k$. In consequence, one can write for $\epsilon_s$ an evolution equation in analogy with $\epsilon$ in the original $k-\epsilon$ model:
\begin{eqnarray}
     \dt{\epsilon}_s = C_{\epsilon_s} \frac{\epsilon_s}{E_s} \dt{E}_s \label{eq:dot_epsilon_s}
\end{eqnarray}
where $C_{\epsilon_s}$ is another model parameter. Developing $\dt{\epsilon}_s = \chi \dt{E}_s + E_s \dt{\chi}$ into Eq. \eqref{eq:dot_epsilon_s} yields:
\begin{eqnarray}
   \dt{\chi} = C_s \frac{\chi}{E_s} \dt{E}_s, 
\end{eqnarray}
where $C_s = C_{\epsilon_s} - 1$. These considerations allow us to write the following closed system:}
\begin{subequations}
    \begin{eqnarray}
        \dt{E}_t &=& F-\epsilon \\
        \dt{\epsilon} &=& C_{k} \frac{\epsilon}{E_t} \dt{E}_t  \left(1+\gamma \frac{F}{\epsilon}\right) - \gamma \dt{F} \\
        \dt{E}_s &=& \left(\left(\frac{\epsilon}{\nu}\right)^{1/2}-\chi\right)E_s \\
        \dt{\chi} &=& C_{s} \frac{\chi}{E_s} \dt{E}_s, \label{eq:dot_chi}
    \end{eqnarray} \label{eq:multiphase-kapibara}
\end{subequations}
where $C_s$ is a third model parameter to be determined. The system Eqs. \eqref{eq:multiphase-kapibara} is one of the main result of the present paper. It consists of four coupled ODEs which allows the kinetic and surface energy to be predicted together with the kinetic and surface energy dissipation/destruction rates. {We just proved that the model equation for $\epsilon_s$ and $\chi$ are formally identical. Therefore, we will hereafter refer to $\chi$ as the destruction rate of surface energy (as we also did previously), although only $\epsilon_s$ can truly (dimensionally) be considered as such.}

{It is important to emphasize that the model for $P$ which is here borrowed to \cite{Vallet1999} is well adapted to characterize the production of surface area due to the tangential strain rate. However, in two-phase flows, the evolution of the interfacial area is also due to topological changes of the fluid structures: it increases with breakup and decreases with coalescence. During break-up, $P$ can potentially reach much larger values than $(\epsilon/\nu)^{1/2}$. Therefore, the model of \cite{Vallet1999} is likely to require corrections in some dense multiphase flows where topological changes can be frequent. The choice we have made here for $P$ should therefore be regarded as a preliminary step}.

{The stretch rate $K$ has a clear physical relevance. It appears as a linear source term in the exact evolution equation for the surface energy, Eq. \eqref{eq:stretch_rate}. Here, the latter is decomposed as $K=P-\chi$. Therefore, the modelling of $\chi$ is inherently linked to the choice of model for $P$. As a result, the quantities $P,~\chi$ (more precisely $F_s$ and $\epsilon_s$) which are here interpreted as a surface production and destruction term, respectively, cannot be viewed as having intrinsic physical significance. Only their difference, $K$, possesses such significance}. {For that reason, the time evolution for $P$ and $\chi$ will not be analysed further in the results section.}

{A final comment concerns the modelling choice adopted for the surface destruction term $\chi$. We adopt here an expression inspired by the $k-\epsilon$ model, an approach that inherently assumes the absence of out-of-equilibrium dynamics. In practical terms, this means we consider that no energy cascade occurs for the surface energy. At present, we lack evidence to confirm the validity of this assumption, so this modelling choice mirrors our current limited understanding of how surface energy is transferred across scales.}

\subsection{Solutions for the linearized system}

Linearizing Eqs. \eqref{eq:multiphase-kapibara} allows the transfer function, i.e. the gain and phase of the different variables with respect to the gain and frequency of the forcing to be predicted. For this purpose, we first write every variable $X \in \{E_k, E_s, E_t, \epsilon, \chi, F\}$ as
\begin{eqnarray}
    X = X^0 + X^\prime = X^0\left[1+\alpha_X \sin \left(\omega t + \phi_X\right)\right]
\end{eqnarray}
where $X^0$ is the steady state value and the prime denotes the perturbation around the base state. The forcing is periodic with a frequency $\omega = 2\pi/T_f$ where $T_f$ is the forcing period. At steady state, in addition to $F^0 = \epsilon^0$, we also have $\chi^0 = \sqrt{\epsilon^0 / \nu}$. The linearized system writes:
\begin{subequations}
    \begin{eqnarray}
        \dt{E_t^\prime} &=& F^\prime - \epsilon^\prime  \\
        \dt{E_s^\prime} &=& A_s^{-1} \left(\frac{1}{2} \chi^0 \epsilon^\prime - \epsilon^0\chi^\prime\right)  ~~~~~~ \textrm{where}~A_s = \frac{\epsilon^0}{E_s^0}=\tau_s^{-1} ~~~[s^{-1}]\\
        \dt{\epsilon^\prime} &=& A_{t} \dt{E_{t}^\prime} - \gamma \dt{F^\prime} ~~~~~~~~~~~~~~~~~~ \textrm{where}~ A_{t} = \frac{C_k(1+\gamma)}{\tau_k + \tau_s}  ~~~[s^{-1}]\\
        \dt{\chi^\prime} &=& A_\chi \frac{\dt{E_s^\prime}}{E_s^0} ~~~~~~~~~~~~~~~~~~~~~~~~~ \textrm{where}~ A_\chi = C_s \chi^0 = C_s {\tau_\eta}^{-1}~~~[s^{-1}]
    \end{eqnarray}
\end{subequations}
Note the appearance of three inverse time-scales, $A_t$, $A_s$ and $A_\chi$ which are related to the kinetic and surface energy time-scales $\tau_{\{k,s\}} = E_{\{k,s\}}^0/\epsilon^0$, and the Kolmogorov time scale $\tau_\eta = (\nu/\epsilon^0)^{1/2}$. We now write perturbations in complex harmonic form $X^\prime = \mathcal{R}\{\widehat{X}e^{i\omega t}\}$. The real part is noted $\mathcal{R}$ and the complex Fourier mode at frequency $\omega$ is  $\widehat{X} e^{i\omega t}$. The hat denotes the complex amplitude, viz $\widehat{X} = |\widehat{X}|e^{i\phi_X}$ with $\phi_X$ being the phase. The linearized system then writes
\begin{subequations}
    \begin{eqnarray}
        i \omega \widehat{E_t} &=& \widehat{F} - \widehat{\epsilon}  \\
        i \omega \widehat{E_s} &=& A_s^{-1} \left(\frac{1}{2} \chi^0 \widehat{\epsilon} - \epsilon^0 \widehat{\chi}\right)  \\
        \widehat{\epsilon} &=& A_{t} \widehat{E_{t}} - \gamma \widehat{F} \\
        \widehat{\chi} &=& A_\chi \frac{\widehat{E_s}}{E_s^0},
    \end{eqnarray}
\end{subequations}
which can be solved to obtain $\widehat{E_t},~\widehat{\epsilon},~\widehat{E_s}$, $\widehat{E_k}$ and $\widehat{\chi}$: 
\begin{subequations}
    \begin{eqnarray}
        \widehat{E_t} &=& \frac{1+\gamma}{A_t(1 + i\omega^*)}\widehat{F} \\
        \widehat{E_s} &=& \frac{1}{2 A_s C_s} \frac{1-\gamma i \omega^*}{(1+i\omega^{\dagger})(1+i\omega^*)} \widehat{F}\\
        \widehat{E_k} &=& \frac{\widehat{F}}{1+i\omega^*} \left(\frac{1+\gamma}{A_t} - \frac{1-\gamma i \omega^*}{2 A_s C_s (1+i\omega^{\dagger})}\right) \\
        \widehat{\epsilon} &=& \frac{1-\gamma i\omega^*}{1+i\omega^*} \widehat{F} \\
        \widehat{\chi} &=& A_\chi \frac{\widehat{E_s}}{E_s^0},
    \end{eqnarray}
\end{subequations}
where $\omega^* = \omega/A_t$ and $\omega^{\dagger} = \omega/A_\chi$. One can then derive the modulus $\alpha_X = |\widehat{X}|/X^0$ :
\begin{subequations}
    \begin{eqnarray}
        \alpha_{E_t} &=& \frac{\alpha_F}{C_k\sqrt{1+(\omega^*)^2}} \\
        \alpha_{E_s} &=& \alpha_\epsilon ~\frac{1}{2C_s} \frac{1}{\sqrt{1+(\omega^{\dagger})^2}} \\
        \alpha_{E_k} &=& \frac{\alpha_{E_t} \alpha_{Es}}{\alpha_\epsilon} \frac{A_k C_k}{A_s A_t} \sqrt{ \big( 2 A_s C_s (1+\gamma) - A_t \big)^2 + \big( 2 A_s C_s (1+\gamma)  \omega^{\dagger} + A_t \gamma \, \omega^* \big)^2 } \\
        \alpha_\epsilon &=& \alpha_{E_t} ~ C_k \sqrt{1+(\gamma \omega^*)^2} \\
        \alpha_{\chi} &=& C_s ~ \alpha_{E_s}
    \end{eqnarray} \label{eq:gain}
\end{subequations}
together with the phase $\phi_X$
\begin{subequations}
    \begin{eqnarray}
        \phi_{E_t} &=& -\arctan(\omega^*) \\
        \phi_{E_s} &=& \phi_\epsilon - \arctan(\omega^{\dagger}) \\
        \phi_{E_k} &=& \phi_{E_t} + \arctan(Z) \\
        \phi_\epsilon &=& \phi_{E_t} - \arctan(\gamma \omega^*) \\
        \phi_\chi &=& \phi_{E_s},
    \end{eqnarray} \label{eq:phase}
\end{subequations}
where 
\begin{equation}
    Z = \frac{\omega^{\dagger} + \gamma \, \omega^{*}}{2 A_s C_s \frac{1 + \gamma}{A_t} \big(1 + (\omega^{\dagger})^2 \big) - \big(1 + \gamma \, \omega^* \, \omega^{\dagger}\big)}.
\end{equation}
We also have $A_k = \tau_k^{-1}$. Eqs. \eqref{eq:gain} and \eqref{eq:phase} provide estimates for the transfer function between $F$ and the other variables. It predicts that, in multiphase turbulence, the evolution of total energy $E_t = E_k + E_s$ and the dissipation rate $\epsilon$ is qualitatively similar to that in the original (single-phase) \textit{Ka-Pi-bara} model \cite{Bos2025,Bos2026}. Note however that the parameters $C_k, \gamma$ and hence $A_t$ can possibly differ between single and multiphase turbulence, resulting in quantitative differences in gain and phase shift. The model thus predicts that in multiphase turbulence, non-equilibrium effects translate into a time delay between the total energy $E_t$ (and possibly $E_k$) and the dissipation rate $\epsilon$. 

The model further shows that $\phi_{E_s} = \phi_\chi$, i.e. surface energy and its destruction are in phase. This is not surprising given that the surface energy is here modelled in analogy with the kinetic energy in the classic $k-\epsilon$ model. It thus predicts that $E_s$ and $\chi$ are in equilibrium, implicitly assuming the absence of a surface energy cascade.

Eqs. \eqref{eq:phase} also predicts the phase shift between the dissipation $\epsilon$ and the surface energy $E_s$. The latter is related to $\omega/A_\chi = \omega \tau_\eta/ C_s$. Since $\tau_\eta$ is increasingly small with increasing Reynolds number, we expect $\omega^{\dagger} \to 0$ when the Reynolds is large. This framework then predicts that $E_s$ and $\epsilon$ tend to be synchronized ($\phi_{E_s} \to \phi_\epsilon$) and proportional to each other ($\alpha_{E_s} \to \alpha_\epsilon / 2 / C_s$) at sufficiently large Reynolds number. The amplitude of $E_s$ is then inversely proportional to $C_s$. If $E_s$ tends to be synchronized with $\epsilon$, then $\chi$ should also be synchronized with $\epsilon$. 

For the kinetic energy, in the limit of large Reynolds numbers $\omega^\dagger \to 0$, we obtain
\begin{equation}
    \lim_{\omega^{\dagger}\to 0} \phi_{E_k} = \phi_{E_t} + \arctan(\gamma \omega^\ddagger) ~~~~ \textrm{where~} \omega^\ddagger = \frac{\omega}{2A_sC_s(1+\gamma)-1},
\end{equation}
which gives a slightly more compact formulation for the gain and phase shift of kinetic energy.

{The linearized solutions for the amplitude and phase shift reveal that the different variables depend on some combined products or sum of the relevant (reciprocal) time-scales $A_t, ~A_s, ~A_\chi$. This is characteristic of a physical system featuring coupled processes, each characterized by their respective physical time-scales. The fact that, for instance, the response for $E_t$ involves a sum of $\tau_k$ and $\tau_s$ indicates that the process is additive. Conversely, if the response depends on a combined product of timescales, as for $E_k$, this implies a multiplicative relationship. }

\section{Numerical database} \label{sec:simulations}

\begin{table}
    \caption{Description of the {DNS} database. To ease readability, it is organised in five distinct sets as described in the leftmost column. The case $Re = 1612, ~We = 12.5, ~\alpha=25.0\%, ~T_f=1.0$ is repeated for clarity.} \label{tab:dns}
\begin{ruledtabular}
    \begin{tabular}{c c c c c c c c c c c c c c c}
    &$Re$ & $We$ & $\alpha$ & $T_f$ & $N$ & $R_\lambda$ & $\tau_k$ & $\tau_s$ & $C_k$ & $\gamma$ & $C_s$ & $A_t$ & $A_s$ & $A_\chi$ \\
    \colrule \\
    single-phase& 645  &  -   &   -  & 1.0 & 128 & 43 & 0.65 &  -   & 0.66 & 0.19 & - & 1.20 & - & - \\
    Reynolds & 1612 & -    & -     & 1.0 & 256 & 72 & 0.70 & - & 0.31 & 0.23 & - & 0.54 & - & - \\
    number & 4166 &    - & -    & 1.0 & 512 & 120 & 0.72 & - & 0.10 & 0.35 & - & 0.19 & - & -  \\
    \\
    multiphase & 1612 & 8$\frac{1}{3}$ & 25\%  & 1.0 & 256 & 60 & 0.58 & 0.36 & 0.78 & 0.20 & 1.48 & 1.00 & 2.76 & 59.54 \\
    Weber & 1612 & 12.5 & 25\%  & 1.0 & 256 & 58 & 0.56 & 0.35 & 0.72 & 0.19 & 1.32 & 0.94 & 2.89 & 53.10\\
    number & 1612 & 25.0 & 25\%  & 1.0 & 256 & 57 & 0.55 & 0.29 & 0.77 & 0.24 & 1.56 & 1.13 & 3.42 & 62.58\\
    \\
    multiphase & 1612 & 12.5 & 12.5\%& 1.0 & 256 & 64 & 0.62 & 0.22 & 0.62 & 0.23 & 2.13 & 0.91 & 4.51 & 85.53\\
    volume & 1612 & 12.5 & 25\%  & 1.0 & 256 & 58 & 0.56 & 0.35 & 0.72 & 0.19 & 1.32 & 0.94 & 2.89 & 53.10\\
    fraction & 1612 & 12.5 & 50\%  & 1.0 & 256 & 52 & 0.50 & 0.43 & 0.84 & 0.20 & 1.12 & 1.08 & 2.33 & 45.16\\
    \\
    multiphase & 645  & 12.5 & 25\% & 1.0 & 128 & 32 & 0.49 & 0.36 & 0.91 & 0.17 & 1.24 & 1.25 & 2.77 & 31.36\\
    Reynolds & 1612 & 12.5 & 25\%  & 1.0 & 256 & 58 & 0.56 & 0.35 & 0.72 & 0.19 & 1.32 & 0.94 & 2.89 & 53.10\\
    number& 4166 & 12.5 & 25\% & 1.0 & 512 & 99  & 0.59 & 0.32 & 0.53 & 0.27 & 1.44 & 0.74 & 3.13 & 92.76 \\
    \\
    multiphase & 1612 & 12.5 & 25\% & 0.75 & 256 & 57 & 0.55 & 0.34 & 0.43 & 0.12 & 1.26 & 0.54 & 2.93 & 50.58\\
    forcing & 1612 & 12.5 & 25\% & 1.0  & 256 & 58 & 0.56 & 0.35 & 0.72 & 0.19 & 1.32 & 0.94 & 2.89 & 53.10\\
    period & 1612 & 12.5 & 25\% & 1.5 &  256 & 57 & 0.55 & 0.35 & 1.26 & 0.36 & 1.58 & 1.92 & 2.88 & 63.39\\
    \\
    \end{tabular}
\end{ruledtabular}
\end{table}

The model is tested against {DNS} of homogeneous isotropic turbulence in a cubic periodic domain. The code \texttt{archer}, the numerical procedure and configuration, have been detailed by \citet{Thiesset2025} and references therein. {These simulations resolve all the scales of the flow without any model.} We use the stochastic forcing of \citet{Eswaran1988} as implemented by \citet{Thiesset2025}, where the amplitude is adjusted at every time step in order to have
\begin{equation}
    F = F^0(1+\alpha_F \sin (\omega t))
\end{equation}
$F_0$ is set to 1 and $\alpha_F$ to {0.5}. No noticeable changes were observed for different $\alpha_F$ (not shown). Note that the forcing of \citet{Eswaran1988} comes with a correlation timescale. We found (not shown) that varying this timescale by a factor 20 did not change the results. The domain width is $L = 1$, which allows us to define a characteristic velocity $U = (F^0L)^{1/3}=1$. We set the same density and same viscosity in both {phases}. The Reynolds is then defined by $Re = UL/\nu \equiv 1/\nu$ and the Weber number is $We = \rho U^2 L/\sigma \equiv 1/\sigma$. Assuming the same density in both {phases} is a rather valid assumption for liquid-liquid emulsions. In addition, it simplifies the analysis as it reduces the parameter space, and it avoids reopening the controversy over whether density jumps give rise to the so-called baropycnal work, which has been claimed to provide an additional contribution to kinetic energy transfer by \citet{Thiesset2025,Narula2026}. 

For all simulations, the viscosity $\nu$ is prescribed in order to guarantee a resolution $\eta/\Delta_x = (\nu^3/F^0)^{1/4} = 1$, where $\eta$ is the Kolmogorov microscale. With $F^0=\epsilon^0=1$ and a domain of size $L=1$ discretized with $N$ mesh cells in each direction, this yields $\nu \equiv N^{-4/3}$. The simulations are run for 40 forcing periods. Four periods are necessary to reach a quasi-limit cycle. We then gather statistics over the next 36 periods (only 30 periods for the two highest Reynolds number cases to be described below). All results are analysed using phase averages. {Statistical convergence for the phase-averaged values was achieved within a 5\% margin. This estimate was obtained by dividing the phase-averaged standard deviation by the square root of the number of forcing periods (either 36 or 30), under the assumption that cycle-to-cycle fluctuations follow a Gaussian distribution.}

Here, we study the effect of the Reynolds and Weber numbers, together with the effect of volume fraction (the volume of a given phase with respect to the entire simulation volume) and of the forcing period. The database is described in Table \ref{tab:dns}. It is organised in five subsets:
\begin{enumerate}
    \item the first three cases correspond to single phase flows at different Reynolds numbers ($Re = 645,~1612,~4166$) varying $\nu = \{15.5,~6.2,~2.4\}\times 10^{-4}$. The forcing period is kept constant $T_f = 1.0$.
    \item The next three cases are for multiphase flows at different Weber numbers, varying $\sigma = 0.04,~0.08,~0.12$ ($We = 25,~12.5,~8\frac{1}{3}$), at constant forcing period $T_f = 1.0$, volume fraction $\alpha = 25\%$ and $Re = 1612$.
    \item The three next cases correspond to different volume fractions $\alpha = 12.5, ~25, ~50\%$, at constant forcing period $T_f = 1.0$, $We = 12.5$ and $Re = 1612$.
    \item The next three cases correspond to different Reynolds numbers, varying $\nu = \{15.5,~6.2,~2.4\}\times 10^{-4}$ ($Re = 645,~1612,~4166$), keeping the forcing period $T_f = 1.0$, the volume fraction $\alpha = 25\%$ and $We = 12.5$ constant.
    \item The last subset employs different forcing period ($T_f = 0.75,~1.0,~1.5$) at constant $\alpha = 25\%$, $We = 12.5$  and $Re = 1612$.
\end{enumerate}

Table \ref{tab:dns} reports the values for $\tau_k = E_k^0/\epsilon^0$, $\tau_s = E_s^0/\epsilon^0$ and $R_\lambda = E_k^0 \sqrt{20/3\nu\epsilon^0}$ (recall $\epsilon^0 = F^0= 1.0$). Note that we refer to $C_k, \gamma, C_s$ as “model parameter” rather than “model constant” since these values are very unlikely to be universal. This reflects the unavoidable trade-off when reducing the full set of multiscale, nonlinear partial differential equations to a simple system of ordinary equations. The objective here is solely to evaluate whether the proposed simplified model can replicate the observed dynamics and to extract qualitative understanding from it. Hence, the model parameters are systematically adjusted for each case using a simple least-square fitting method comparing the phase-averaged {DNS} data to the linearized solutions. The procedure minimizes a global residual constructed from the normalized errors in $E_k, ~E_s, ~\dt{E}_k, ~\dt{E}_s, ~\epsilon$ and $\chi$. {The $L^2$-norm of this residual after the optimization process is generally below 10\%, except for some rare pathological cases where it can reach 15\%. This gives the overall model accuracy.} The values for the model parameters $C_k, \gamma, C_s$ together with $A_t, ~A_s, ~A_\chi$ appearing in the linearized solutions are reported in Table \ref{tab:dns}. 

Table \ref{tab:dns} reveals that the values for $C_k, ~\gamma, ~C_s$ are clearly not constant. In contrast, some trends in their evolution with respect to the flow parameters are discernible. It appears that $C_k$ systematically decreases when the Reynolds number increases. It is also larger in multiphase than in single phase flows, it increases with $\alpha$ but does not seem to vary with $We$. The parameter $\gamma$ increases with $Re$ and is slightly smaller in multiphase turbulence compared to the single phase case. For the newly introduced parameter $C_s$ which comes out from the evolution equation for $\chi$, it increases with $Re$ and decreases with $\alpha$. 

{The variability of the parameters $ C_k, ~\protect \gamma, ~C_s$ reflects the absence of certain physical processes in the model. Recall that the \textit{KaPiBara} model is derived from considerations at infinite Reynolds numbers, while we investigate here turbulent flows at finite Reynolds numbers. As a result, the variations observed in the constants in multiphase flows may also be observed in single-phase turbulence at comparable Reynolds numbers. For instance, the observed variations of the constants $C_k,~\protect \gamma$ with $T_f$ in multiphase flows were observed to happen in similar proportions in the single-phase case (not shown). The sensitivity to the Reynolds number can possibly be leveraged to refine the model, as demonstrated by \citet{Bos2026}, who proposes a finite Reynolds number correction. This modification introduces an additional constant. Other limitations of the model, as well as potential improvements, will be addressed in the results section and further discussed in the conclusions.}

From the linearized solution of the system, it follows that the relevant quantities for predicting the dynamics are $A_t, ~A_s, ~A_\chi$. Table \ref{tab:dns} first indicates that for most cases, $A_\chi \gg 1$ suggesting that the assumption $\omega^{\dagger} \to 0$ is rather well justified. Hence, we expect $\epsilon$ and $E_s$ to be in phase or close to it. We then should have that $\epsilon$ and $\dt{E}_s$ are lagged by $\pi/2$, i.e. the conversion is maximum when kinetic (or total) energy dissipation fluctuation is zero. The inverse of the characteristic timescale for the total energy $A_t$ decreases quite substantially when $Re$ increases but does not seem to vary much with the other flow parameter $\alpha, ~We$. What is more surprising is the evolution of $C_k, \gamma$ together with $A_t$ with the forcing period as we would have expected these values to be intrinsic quantities of the flow, but not of the forcing. An explanation for this trend is proposed at the end of the results section \S \ref{sec:results}. 

\begin{figure}
    \begin{tabular}{p{0.49\textwidth}p{0.49\textwidth}}
        ~~~~~~(a) & ~~~~(b)
    \end{tabular}
    \includegraphics[width=\textwidth]{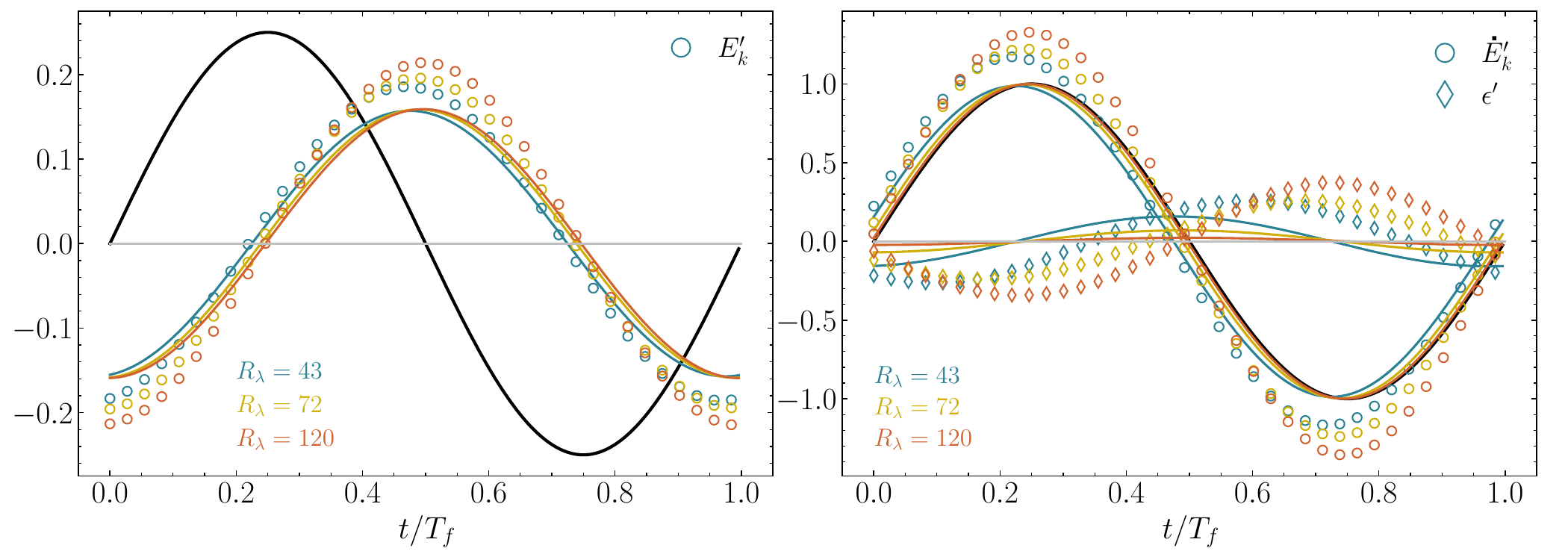} \\
    \begin{tabular}{p{0.49\textwidth}p{0.49\textwidth}}
        ~~~~~~(c) & ~~~~(d)
    \end{tabular}
    \includegraphics[width=\textwidth]{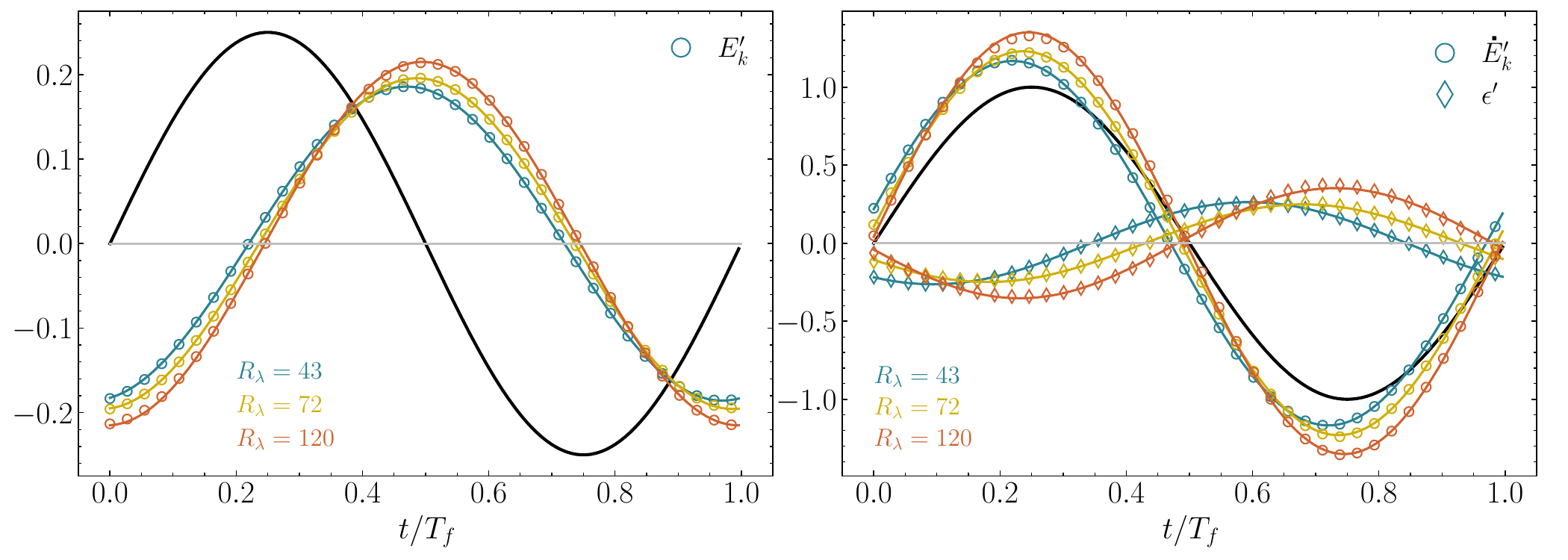}
    \caption{Predictions using the $k$--$\epsilon$ model (top panel, (a, b)) and the \textit{Ka-Pi-bara} model (bottom panel, (c, d)) in single phase flows. {The symbols represent the {DNS} while the lines correspond to the predictions using the linearized model. In (a) and (c), $\mcirc$ represents $E_k^\prime$ while in (b) and (d) $\mdiamond$ is for $\epsilon^\prime$, and $\mcirc$ for $\protect \dt{E}_k^\prime$. The different colours \{blue, yellow, red\} represent different Reynolds numbers $Re = 645, ~1612, ~4166$, respectively, corresponding to $R_\lambda = 43,~72,~120$, as indicated in the legend. The black curve in (a-d) represents $F^\prime$}. The kinetic energy $E_k$ is normalised by $F^0 \alpha_F T_f$ while the terms of its balance equation are normalised by $F^0 \alpha_F$.} \label{fig:single_phase}
\end{figure}

\section{Results} \label{sec:results}

\subsection{Single-phase flows}

We first analyse the results obtained with the $k$--$\epsilon$ and \textit{Ka-Pi-bara} models for single-phase flows. Figure~\ref{fig:single_phase} presents the results for $Re = 645,~1612,~4166$, corresponding to $R_\lambda = 43,~72,~120$. Panels (a, c) show the phase-averaged kinetic energy $E_k^\prime$, while panels (b, d) display the different terms of the phase-averaged kinetic energy budget, $F^\prime,~\dt{E}^\prime_k,~\epsilon^\prime$. The simulation results are indicated by symbols. {The curves correspond to the solutions of linearized system. These were observed to be indistinguishable from solutions of the non-linear model.} This means that, within the investigated regimes, non-linearities have a limited impact. {Therefore, in the figures presented hereafter, only the predictions of the linearized system will be displayed.}

Results from {DNS} indicate that $E_k^\prime$ and $\epsilon^\prime$ are increasingly lagged when $Re$ increases. This behaviour, known as non-equilibrium effects cannot be reproduced by the $k$--$\epsilon$ model (Figure~\ref{fig:single_phase}(a, b)) which predicts that both quantities are in phase. As a consequence, the departures between observations and the classical $k$--$\epsilon$ model are quite significant and increase with $Re$. In particular, the predicted amplitude for the fluctuations of $\epsilon$ are drastically underestimated. 

In contrast, the \textit{Ka-Pi-bara} model accurately reproduces the phase lag between $E_k$ and $\epsilon$ which is encoded in the parameter $\gamma$. The latter increases with $Re$ (see Table \ref{tab:dns}) consistently with observations. Irrespective of the Reynolds number, the predictions agree very nicely with {DNS}. Therefore, in what follows, only the results of the \textit{Ka-Pi-bara} model will be discussed.

\subsection{Multiphase flows}

\subsubsection{Effect of Weber number} 

\begin{figure}
    \begin{tabular}{p{0.49\textwidth}p{0.49\textwidth}}
        ~~~~~~(a) & ~~~~(b)
    \end{tabular}
    \includegraphics[width=\textwidth]{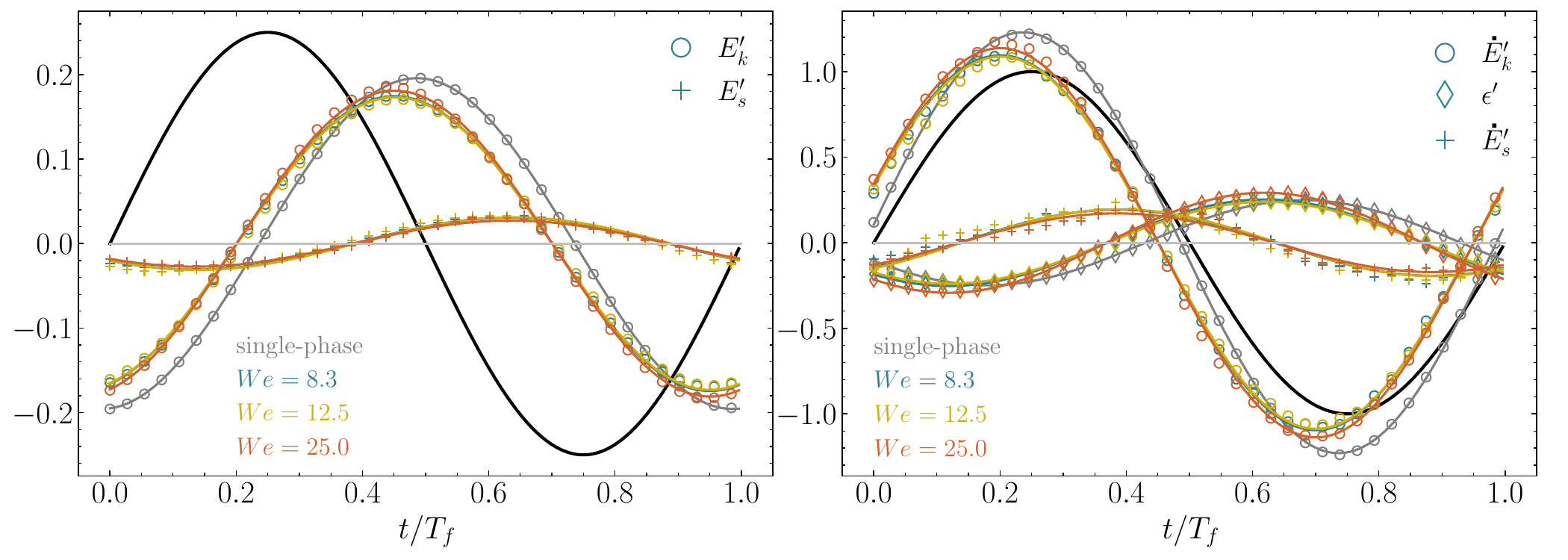} 
    \caption{Effect of the Weber number. {The symbols represent {DNS} while the lines correspond to the predictions using the linearized model. In (a), $\mcirc \equiv E_k^\prime$ and $\mplus$ is for $E_s^\prime$. In (b), $\mdiamond$ is for $\epsilon^\prime$, $\mcirc$ is for $\protect \dt{E}_k^\prime$ and $\mplus$ is for $\protect \dt{E}_s^\prime$. Data represented in gray correspond to the single-phase case. The other colours \{blue, yellow, red\} represent different Weber numbers $We = 8.33, ~12.5, ~25$, respectively, as indicated in the legend. The black curve in (a) and (b) represents $F^\prime$}. Energies $E_k$ and $E_s$ are normalised by $F^0 \alpha_F T_f$ while the terms of the balance equation are normalised by $F^0 \alpha_F$.}\label{fig:multiphase_Weber}

\end{figure}

We now proceed with the analysis of multiphase turbulence. We start by exploring the effect of Weber number in Fig. \ref{fig:multiphase_Weber}. In the range of $We$ investigated, the surface energy reservoir $E_s^0$ represent about roughly 60\% of the kinetic energy reservoir (see Table \ref{tab:dns}), i.e. nearly 35\% of the total energy. Interestingly, this proportion does not vary much with the Weber number. Since $E_s^0 = \sigma A^0$, this means that decreasing the surface tension is compensated by an increase of the surface area density $A^0$ so that the surface energy is constant. 

Similarly, phase-averaged fluctuations of surface energy $E_s^\prime$ and $E_k^\prime$ (Fig. \ref{fig:multiphase_Weber}(a)) do not seem to be influenced by the Weber number. All curves collapse rather nicely, and the same observation applies to the different terms of the energy balance (Fig. \ref{fig:multiphase_Weber}(b)). The model accurately reproduces observations. In particular, we find that $E_s^\prime$ and $\epsilon^\prime$ are almost in phase as anticipated by the analysis of the linearized system for $\omega^\dagger \to 0$. 

Overall, the results indicate that fluctuations of $E_k^\prime$ precedes fluctuations of $E_s^\prime$ which is more dynamically related to $\epsilon^\prime$. Comparing these results to single-phase flows at same $Re$ indicates that the peak of kinetic energy and dissipation in multiphase flows appears earlier than in single-phase flows. This is a consequence of the kinetic energy being converted into the surface energy reservoir. 

It is often stated that the presence of an interface increases the amount of kinetic energy dissipation. Here, we note that the presence of the interface does not alter much the amplitude of the phase-averaged dissipation which appears to be roughly the same in single-phase or multiphase flows. This observation is likely to break down when the two phases have significantly different densities \cite[see e.g.][]{Dodd2016,Thiesset2025}.

Again, {although results are not presented here, we were not able to distinguish between the non-linear and the linearized solutions}. This suggests again that non-linearities do not play too much role. Both predictions show very good agreement with {DNS} as they correctly capture the dynamical behaviour of energy conversion in multiphase flows. Note again the advantage of using the \textit{Ka-Pi-Bara} model compared to the classical $k-\epsilon$ model. Indeed, without the introduction of the non-equilibrium constant $\gamma$, the phase shift between $E_k^\prime$ and $\epsilon^\prime$ would not have been captured and hence, the dynamics of $E_s^\prime$ would not have been correctly represented.

\subsubsection{Effect of volume fraction} 

\begin{figure}

    \begin{tabular}{p{0.49\textwidth}p{0.49\textwidth}}
        ~~~~~~(a) & ~~~~(b)
    \end{tabular}
    \includegraphics[width=\textwidth]{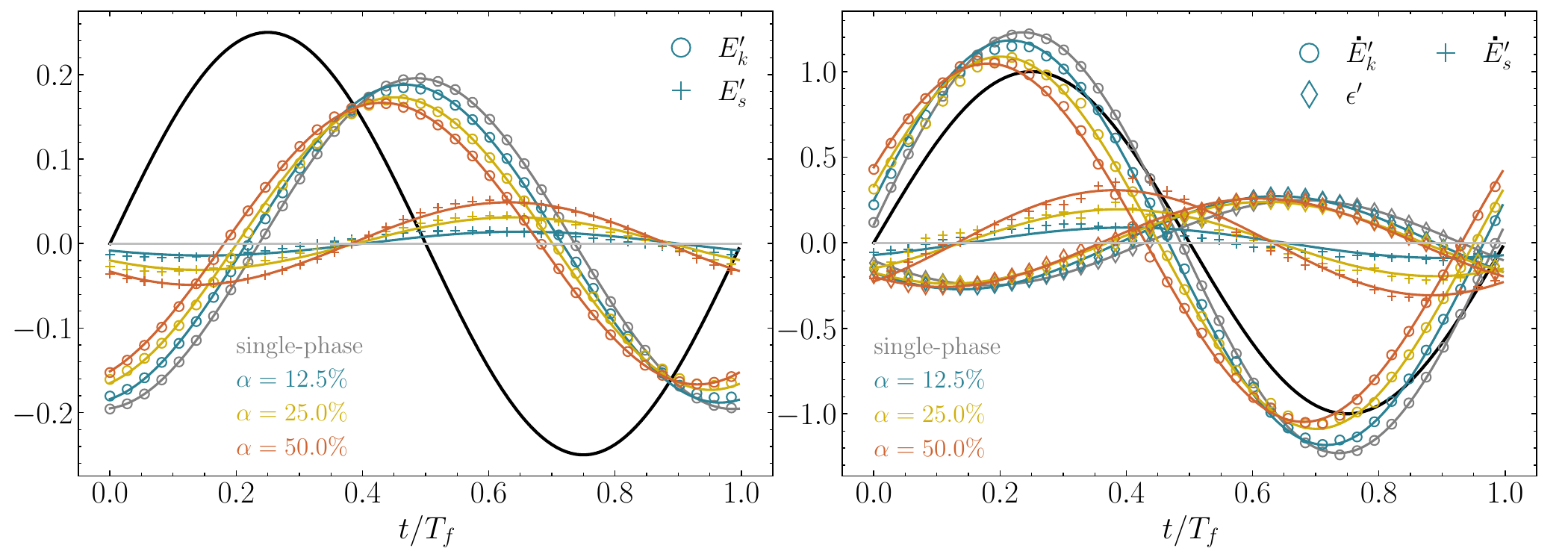}
    \caption{Effect of the volume fraction. {The symbols represent the {DNS} while the lines correspond to the predictions using the linearized model. In (a), $\mcirc$ corresponds to $E_k^\prime$ and $\mplus$ to $E_s^\prime$. In (b) $\mdiamond$ is for  $\epsilon^\prime$, $\mcirc$ is $\protect \dt{E}_k^\prime$ and $\mplus$ is $\protect \dt{E}_s^\prime$. Data represented in gray correspond to the single-phase case. The other colours \{blue, yellow, red\} represent different volume fractions $\alpha = 12.5, ~25.0, ~50.0\%$, respectively, as indicated in the legend. The black curves in (a) and (b) represents $F^\prime$}. Energies $E_k$ and $E_s$ are normalised by $F^0 \alpha_F T_f$ while the terms of the balance equation are normalised by $F^0 \alpha_F$.} \label{fig:multiphase_alpha}
\end{figure}

In contrast with variations of the Weber number, changes in $\alpha$ lead to different proportions between kinetic and surface energy in the system. Table \ref{tab:dns} indicates that $E_s^0$ is multiplied by roughly 2 between $\alpha = 12.5\%$ and $\alpha = 50\%$ while $E_k^0$ is slightly decreasing. At $\alpha = 50\%$, $E_k^0$ and $E_s^0$ appear in comparable proportion. 

Similarly, when $\alpha$ increases, the amplitude of the phase-averaged surface energy $E_s^\prime$ increases, while the amplitude of $E_k^\prime$ decreases (Fig. \ref{fig:multiphase_alpha}(a)). The amplitude of $E_s^\prime$ is encoded in the parameter $C_s$ whose value drives the amount of surface energy destruction $\chi$, with respect to surface energy production $P=(\epsilon/\nu)^{1/2}$. Table \ref{tab:dns} indicates that the parameter $C_s$ decreases by almost a factor 2 between $\alpha=12.5$ and 50\% which corresponds to roughly the same ratio on the amplitude of $E_s^\prime$ on Fig. \ref{fig:multiphase_alpha}(a). Speculatively, $C_s$ could exhibit greater universality when parameterized by $\alpha$ or $E_s^0/E_k^0$.

It is observed that the phase shift for the surface energy signals does not depend on $\alpha$ (Fig. \ref{fig:multiphase_alpha}(a)). This behaviour is anticipated by the model, which predicts that the peak should coincide with that of $\epsilon^\prime$ (recall that $A_\chi \gg 1$). This is confirmed by observing the evolution of $\epsilon^\prime$ (Fig. \ref{fig:multiphase_alpha}(b)) which does not vary much, either in amplitude or in phase, with $\alpha$. However, the kinetic energy signals $E_k^\prime$ are systematically shifted toward earlier times as the volume fraction increases. This indicates that when a larger amount of surface energy is present into the system, the conversion of kinetic energy into surface energy is obviously enhanced and as a consequence, kinetic energy reaches its peak earlier and with a lower amplitude. A contrario, the signals for small $\alpha$ approaches the one corresponding to single phase flows. The influence of the interface and energy conversion processes on the energy balance might become negligible for $\alpha<10\%$.

\subsubsection{Effect of Reynolds number}

We now explore the effect of Reynolds number, while the volume fraction $\alpha$, Weber number and forcing period are kept constant. Table \ref{tab:dns} indicates that the Reynolds number does not influence much the proportion between $E_s^0$ and $E_k^0$, their ratio remains roughly constant at a value of about 60\%. 

\begin{figure}
    \begin{tabular}{p{0.49\textwidth}p{0.49\textwidth}}
        ~~~~~~(a) & ~~~~(b)
    \end{tabular}
    \includegraphics[width=\textwidth]{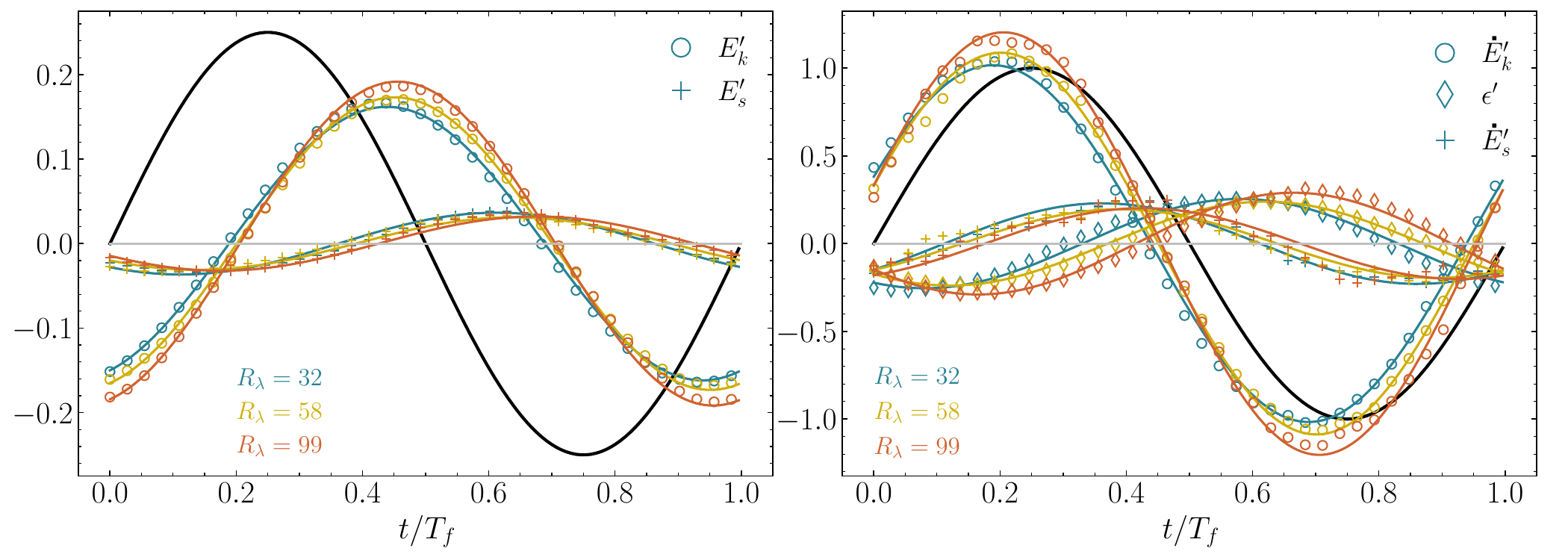}
    \caption{Effect of the Reynolds number. {The symbols represent the {DNS} while the lines correspond to the predictions using the linearized model. In (a), $\mcirc$ corresponds to $E_k^\prime$ and $\mplus$ to $E_s^\prime$. In (b) $\mdiamond$ is for  $\epsilon^\prime$, $\mcirc$ is $\protect \dt{E}_k^\prime$ and $\mplus$ is $\protect \dt{E}_s^\prime$. The colours \{blue, yellow, red\} represent different Reynolds numbers $Re = 645, ~1612, ~4166$, respectively, corresponding to $R_\lambda = 32,~58,~99$ as indicated in the legend. The black curve in (a) and (b) represents $F^\prime$}. Energies $E_k$ and $E_s$ are normalised by $F^0 \alpha_F T_f$ while the terms of the balance equation are normalised by $F^0 \alpha_F$.}\label{fig:multiphase_Reynolds}
\end{figure}

As already observed in single-phase flows, non-equilibrium effects, i.e. the phase lag between $E_k^\prime$ (fig. \ref{fig:multiphase_Reynolds}(a)) and $\epsilon^\prime$ (fig. \ref{fig:multiphase_Reynolds}(b)) become more pronounced as $Re$ increases. This is nicely reproduced by the \textit{Ka-Pi-bara} model as the non-equilibrium parameter $\gamma/A_t$ increases by almost a factor 3 between the two extreme Reynolds number values (see Table \ref{tab:dns}). 

A careful examination of Figure \ref{fig:multiphase_Reynolds} indicates that a phase lag between $E_s^\prime$ and $\epsilon^\prime$ is perceptible. The latter is substantial at the lowest Reynolds number but becomes negligible as $Re$ increases. This trend is nicely reproduced by the model through the parameter $C_\chi \sim \tau_\eta^{-1}$ which increases in such a way that $\omega^\dagger \to 0$ when $Re$ increases (see Table \ref{tab:dns}). With appropriate values for the parameters $C_k, \gamma, C_s$, the model captures nicely the observed dynamics of the different variables. For instance, it reproduces the slight increase of kinetic energy $E_k^\prime$ with $Re$, as was also observed in single-phase flows (see Fig. \ref{fig:single_phase}). It also predicts correctly that the amplitude of $E_s^\prime$ does not vary much with $Re$ and the same applies to $\epsilon^\prime$.

{In the study by \citet{Mukherjee2019}, the phase shifts were analysed by computing the cross-correlations between several variables. Only one case (named T5) is reported in their work, characterized by the following flow parameters: $R_\lambda = 90$, $\alpha=0.1$ and $We \approx 5$. They observed a phase shift of $0.23\tau_k$ between $E_k$ and $\epsilon$ (more precisely the enstrophy $\langle |\vect{\omega}|^2 \rangle = \epsilon / \nu$), where $\tau_k = E_k^0/\epsilon^0$ is the present definition for the eddy turnover time (not to be confused with $\tau_k$ in \cite{Mukherjee2019} which represents the Kolmogorov timescale). }

{In our database, the closest point to T5 of \citet{Mukherjee2019} is $R_\lambda = 99$, $\alpha=0.25$ and $We=12.5$. For this case, we find a phase shift between $E_k$ and $\epsilon$ of roughly 0.2$T_f$ which corresponds to about $0.35 \tau_k$, i.e. a slightly larger value. This difference can be explained by the difference in volume fraction: $\alpha=0.1$ in \citet{Mukherjee2019}, and $\alpha=0.25$ here. The results of Fig. \ref{fig:multiphase_alpha} reveal that the larger the volume fraction, the larger the phase shift between $E_k$ and $\epsilon$. Hence, our findings are consistent with the work by \citet{Mukherjee2019}, although this comparison remains qualitative.}

\subsubsection{Effect of forcing period}

\begin{figure}
    \begin{tabular}{p{0.49\textwidth}p{0.49\textwidth}}
        ~~~~~~(a) & ~~~~(b)
    \end{tabular}
    \includegraphics[width=\textwidth]{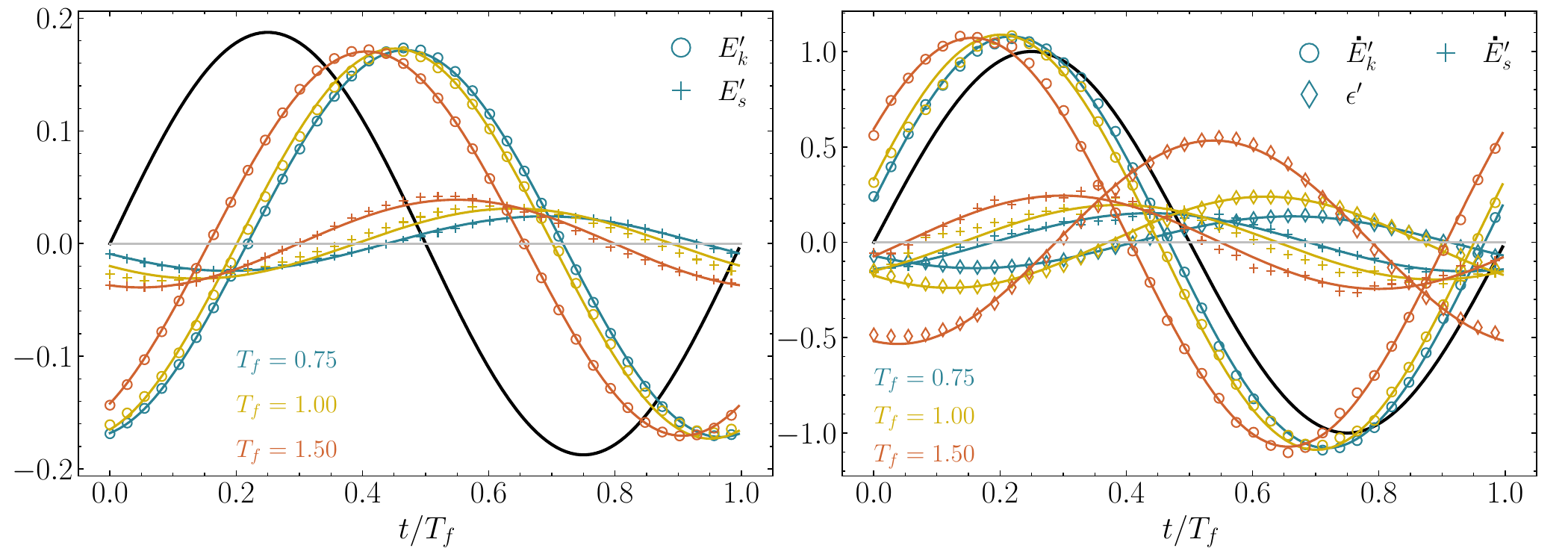} 
    \caption{Effect of the forcing period. {The symbols represent the {DNS} while the lines correspond to the predictions using the linearized model. In (a), $\mcirc$ corresponds to $E_k^\prime$ and $\mplus$ to $E_s^\prime$. In (b) $\mdiamond$ is for  $\epsilon^\prime$, $\mcirc$ is $\protect \dt{E}_k^\prime$ and $\mplus$ is $\protect \dt{E}_s^\prime$. The colours \{blue, yellow, red\} represent different forcing periods $T_f = 0.75, ~1.0, ~1.5$, respectively, as indicated in the legend. The black curve in (a) and (b) represents $F^\prime$}. Energies $E_k$ and $E_s$ are normalised by $F^0 \alpha_F T_f$ while the terms of the balance equation are normalised by $F^0 \alpha_F$.}\label{fig:multiphase_Tf}
\end{figure}

We now vary the forcing period $T_f$. The other parameters are kept constant so that the proportion between surface and kinetic energy remains the same (see Table \ref{tab:dns}).

Fig. \ref{fig:multiphase_Tf}(a,b) indicates that when $T_f$ decreases (increasing $\omega$), the amplitude of $\epsilon^\prime$ decreases. We can thus expect $\dt{E}_k^\prime \to F^\prime$ in the limit of small $T_f$ (large $\omega$), i.e. in the frozen regime. A contrario, the amplitude of $\epsilon^\prime$ continuously increases as $T_f$ increases. Hence, the quasi steady-state limit $\epsilon^\prime \to F^\prime$ is likely to be recovered for large $T_f$ (small $\omega$). In this respect, the dynamical behaviour of periodically multiphase turbulence is qualitatively similar to the single-phase case.

The amplitude of $E_s^\prime$ does not increase in same proportion as $\epsilon^\prime$. This translates in a substantial increase of the parameter $C_s$ when $T_f$ increases (see Table \ref{tab:dns}). Irrespective of $T_f$, the value of $A_\chi$ remains sufficiently large for $\epsilon^\prime$ and $E_s^\prime$ to evolve almost in phase. Again, this shows that the dynamics of surface energy is coupled to the dynamics of the kinetic energy dissipation rate $\epsilon$. 

Once the parameters $C_k, ~\gamma, ~C_s$ adjusted, the model reproduces the dynamics of each variable fairly well. It is quite surprising however that these parameters change with the forcing frequency. It would have been much more natural to expect these parameters to be intrinsic to the flow with no relation with the properties of the external energy input. To explain this behaviour, we should recall that the \textit{Ka-Pi-bara} model was derived from a two-scale decomposition, distinguishing, on the one hand, the large scales whose dynamics are driven by the forcing and energy transfer and, on the other hand, the small-scales where only the energy transfer and dissipation operates. This effectively results in the appearance of two characteristic timescales, i.e. $A_t^{-1}$, which appears in the expression for $\widehat{E_k}$ (the large scales), and $\gamma A_t^{-1}$ which appears in the expression for $\widehat{\epsilon}$ (the small-scales). But clearly, these two timescales are not independent, i.e. the timescale of the small-scales is $\gamma$ times the timescale of the large scales. This is likely to cause $C_k, \gamma$ (and eventually $C_s$) to adapt in order to reproduce the correct dynamics. One could have anticipated this since the \textit{Ka-Pi-bara} model was derived hypothesizing that the small-scales rapidly (=instantaneously) adapt to the large scales (see hypothesis (1) in \cite{Bos2025}), i.e. $E_k^> = \gamma E_k^<$. This suggests that the performance of the \textit{Ka-Pi-bara} can be further improved by introducing a two-timescale decomposition alongside the two-scale decomposition. This will unfortunately result in additional ODEs and hence additional model parameters.

\section{Conclusions} \label{sec:conclusions}

This paper examines how conversions between kinetic and surface energy influence the overall energy balance of multiphase turbulence. A time-periodic forcing is imposed to assess the cycle connecting energy injection, conversion, and dissipation. We use phase-averages to analyse data from {direct} numerical simulations spanning different Reynolds and Weber numbers, volume fractions, and forcing frequencies. {In parallel}, a model, viewed as a generalization of the $k$--$\epsilon$ model to non-equilibrium multiphase turbulence, is proposed to reproduce and explain the observed dynamics.

To derive this model, we first propose to reformulate the \textit{Ka-Pi-bara} model in terms of $E_t = E_s + E_k$, exploiting the analogy between the balance equation for $E_t$ in multiphase flows and the kinetic energy in single-phase flows. The evolution of surface energy is decomposed into a production term, related to the Kolmogorov strain rate, and a destruction term. Due to limited knowledge of the latter, we introduce an evolution equation, written in analogy with the kinetic energy dissipation rate in the standard $k$--$\epsilon$ model. The proposed system then consists of four coupled ODEs for $E_t, ~\epsilon, ~E_s, ~\chi$, respectively, featuring three parameters $C_k, ~\gamma, ~C_s$. The proposed model is then linearized in order to highlight the relevant timescales of the system and to get predictions of how the different variables respond to the energy input.

The present work yields some outcomes that are listed below. 

We find that solutions of the linearized and non-linear systems do not reveal substantial differences, suggesting that non-linearities are weak. Furthermore, once the model parameters adjusted, the predictions agree satisfactorily with {DNS}. Trends in the evolution of optimal values for $C_k, ~\gamma, ~C_s$ with respect to the control parameters indicate their non-universality. Nonetheless, the dependence of $C_k$ and $\gamma$ to $Re$ and $\alpha$ appears potentially parametrisable, as does that of $C_s$ with respect to $\alpha$. {We have found that $C_s$ could potentially be recast as a function of $\alpha(1-\alpha)$, to reveal greater universality.} If such parametrizations were to be confirmed, some more general predictions for the energy balance in multiphase turbulence could be obtained, with potential applications in various engineering problems. 

The linearized solutions indicate that fluctuations of $E_s$ and its destruction rate are synchronized. {This model thus implicitly assumes that} the surface energy is in equilibrium, underscoring the absence of a surface energy cascade. $E_s$ and $\chi$ further appear to be synchronized with the kinetic energy dissipation at sufficiently large $Re$. However, there is a substantial time lag between $E_k$ (or $E_t$) and its dissipation $\epsilon$, a key feature of non-equilibrium turbulence that is well captured by the \textit{Ka-Pi-Bara} model.  

{Despite the adopted hypothesis, the model reproduces the observed dynamics, albeit with three model parameters adjusted. While this result does not constitute definitive proof, it suggests that both the chosen formulation for $P$ and the assumption of the absence of a surface energy cascade represent reasonable approximations within the studied regimes. Future research is now necessary to rigorously validate these assumptions. Achieving this will require advancing our understanding of surface energy transfer mechanisms across scales.}

A systematic exploration of the parameter space shows that the Weber number does not significantly affect the base flow state or the fluctuations about it. This implies a compensation between surface tension and surface area. The effect of volume fraction is much more perceptible. When the two phases are in equal proportion $\alpha = 50\%$, fluctuations associated with the surface energy are maximum. For iso-density fluid flows, the influence of an interface on the energy balance might become negligible for $\alpha < 10\%$. An increase in Reynolds number leads to a larger phase lag between $E_k$ and $\epsilon$, indicating a stronger influence of non-equilibrium effects due to the increased separation between large and small scales. The surface energy is much less influenced by the effect of Reynolds number, {hinting at a weaker or absent cascade for the surface energy}. Varying the forcing frequency reveals that the model can still adapt, albeit at the cost of adjusting the parameters $C_k, ~\gamma, ~C_s$. {The model is based on the assumption that $\protect \gamma$ does not vary in time. It may thus lack a timescale decomposition, which reflects that the small scales do not instantaneously adapt to the large scales. The precise time evolution for $\gamma$ could be derived from the time evolution of the scale-by-scale energy distribution for the base state ${E_k^>}^0, {E_k^<}^0$ and perturbations ${E_k^>}^\prime, {E_k^<}^\prime$, as was proposed by e.g. \citet{Yoshizawa1994, Bos2024}. This opens directions for further improvements of the model, with the hope it would lead to some more universal constants. }

\acknowledgments
The authors acknowledge the support of the French National Research Agency (ANR) through the project ANR--22-CE51-0005 (DeltaPhi). Calculations were performed using the computing resources of CRIANN (Normandy, France), under the project 2018002. We also benefited from discussions with W. Bos and R. Zamansky which are gratefully acknowledged. Members of the archer team, namely J.-C. Brändle de Motta, B. Duret, T. Ménard and A. Poux are also acknowledged.

\bibliography{multiphase_KapiBara.bbl}
\end{document}